\documentclass[twocolumn]{aastex62}
\graphicspath{{/Volumes/Maxtor/subbands/},{/media/alberto/TRANSCEND/UDS/}}
\usepackage{comment}
\usepackage{multirow}
\usepackage{array}
\usepackage{xspace}

\newcommand{\xmm}{XMM-\textit{Newton}\xspace}
\newcommand{\chandra}{\textit{Chandra}\xspace}
\newcommand{\swiftbat}{\textit{Swift}-BAT\xspace}
\newcommand{\nustar}{\textit{NuSTAR}\xspace}
\newcommand{\nh}{$N_{\rm H}$}
\newcommand{\nhcgs}{cm$^{-2}$}
\newcommand{\fluxcgs}{erg cm$^{-2}$ s$^{-1}$\xspace}

\newcommand{\cid}{\texttt{cid\_3570}\xspace}
\shortauthors{A. Masini et al.}

\begin{document}
\title{THE \nustar EXTRAGALACTIC SURVEYS: UNVEILING RARE, BURIED AGNS AND DETECTING THE CONTRIBUTORS TO THE PEAK OF THE COSMIC X-RAY BACKGROUND}

\correspondingauthor{Alberto Masini}
\email{alberto.masini@dartmouth.edu}

\author[0000-0002-7100-9366]{Alberto Masini}
\affil{Department of Physics and Astronomy, Dartmouth College, 6127 Wilder Laboratory, Hanover, NH 03755, USA}
\affil{INAF-Osservatorio di Astrofisica e Scienza dello Spazio di Bologna, via Gobetti 93/3, 40129 Bologna, Italy}
\affil{Dipartimento di Fisica e Astronomia (DIFA), Universit\`a  di Bologna,  via Gobetti 93/2, 40129 Bologna, Italy}

\author{Andrea Comastri}
\affil{INAF-Osservatorio di Astrofisica e Scienza dello Spazio di Bologna, via Gobetti 93/3, 40129 Bologna, Italy}

\author{Francesca Civano}
\affil{Harvard-Smithsonian  Center  for  Astrophysics,  60 Garden Street, Cambridge, MA 02138, USA}

\author{Ryan C. Hickox}
\affil{Department of Physics and Astronomy, Dartmouth College, 6127 Wilder Laboratory, Hanover, NH 03755, USA}

\author{Christopher M. Carroll}
\affil{Department of Physics and Astronomy, Dartmouth College, 6127 Wilder Laboratory, Hanover, NH 03755, USA}

\author{Hyewon Suh}
\affil{Subaru Telescope, National Astronomical Observatory of Japan, 650 A'ohoku place, Hilo, HI, 96720, USA}

\author{William N. Brandt}
\affil{Department of Astronomy and Astrophysics, 525 Davey Lab, The Pennsylvania State University, University Park, PA 16802, USA}
\affil{Institute for Gravitation and the Cosmos, The Pennsylvania State University, University Park, PA 16802, USA}
\affil{Department of Physics, 104 Davey Laboratory, The Pennsylvania State University, University Park, PA 16802, USA}

\author{Michael A. DiPompeo}
\affil{Department of Physics and Astronomy, Dartmouth College, 6127 Wilder Laboratory, Hanover, NH 03755, USA}

\author{Fiona A. Harrison}
\affil{Cahill Center for Astronomy and Astrophysics, California Institute of Technology, Pasadena, CA 91125, USA}

\author{Daniel Stern}
\affil{Jet Propulsion Laboratory, California Institute of Technology, 4800 Oak Grove Drive, Pasadena, CA 91109, USA}

\begin{abstract}

We report on the results of active galactic nuclei (AGNs) detection by \nustar performed in three extragalactic survey fields (COSMOS, UDS, ECDFS) in three hard bands, namely H1 ($8-16$ keV), H2 ($16-24$ keV) and VH ($35-55$ keV). The aggregated area of the surveys is $\sim$ 2.7 deg$^2$. While a large number of sources is detected in the H1 band (72 at the 97\% level of reliability), the H2 band directly probing close to the peak of the Cosmic X-ray Background (CXB) returns four significant detections, and two tentative, although not significant, detections are found in the VH band. All the sources detected above 16 keV are also detected at lower energies. We compute the integral number counts for sources in such bands, which show broad consistency with population synthesis models of the CXB. We furthermore identify two Compton-thick AGNs, one in the COSMOS field, associated with a hard and faint \textit{Chandra} source, and one in the UDS field, never detected in the X-ray band before. Both sources are at the same redshift $z\sim1.25$, which shifts their Compton-hump into the H1 band, and were previously missed in the usually employed \nustar bands, confirming the potential of using the H1 band to discover obscured AGNs at $z>1$ in deep surveys.

\end{abstract}

\section{Introduction} \label{sec:intro}

X-ray surveys are one of the most effective ways to detect, select and identify accreting supermassive black holes \citep[see][for a review]{brandtalexander15}. In the past decades, comprehensive X-ray surveys by \xmm and \chandra covered a wide range in the flux-area plane, exploring a large range in redshift and luminosity, and characterizing the properties and evolution of active galactic nuclei \citep[AGNs;][]{cappelluti09,elvis09,xue11,ranalli13,luo17}, which are the major contributors to the diffuse extragalactic emission named the Cosmic X-ray Background (CXB). The intensity of the CXB can be ascribed to a mixed contribution from all different kinds of AGNs \citep[e.g.,][]{settiwoltjer89,comastri95}, showing a large range in obscuration, luminosity and redshift. In particular, a non-negligible contribution from a class of heavily obscured AGNs, called Compton-thick (\nh$>10^{24}$ \nhcgs), is required in order to successfully explain the intensity of the CXB around its peak at $\sim20-30$ keV \citep[e.g.,][]{gilli07, treister09, draperballantyne10, akylas12}.
\par Despite being very successful in detecting and describing the mixture of such different populations at low energies ($<$ 10 keV), X-ray surveys are affected by a substantial absorption bias, mainly in the local universe ($z<1$), where most of the effects of gas obscuration along the line of sight are seen. Depending on the degree of obscuration, the intrinsic flux of a source can be significantly attenuated up to $\sim 10-20$ keV, ultimately driving the source to be undetected in deep surveys. The effects of such obscuration become less significant at high energies, in particular in the hard X-ray band ($E>10$ keV). However, hard X-ray surveys performed in the past years with coded-mask instruments like INTEGRAL and \swiftbat detected a tiny fraction of the obscured sources making up the majority of the CXB above 10 keV \citep{krivonos05,ajello08}. 
\par With the advent of \nustar \citep{harrison13}, the first focusing hard X-ray telescope (comprising two focal plane modules, namely FPMA and FPMB), sensitive hard X-ray surveys above 10 keV started to be feasible, allowing sources to be detected at $\sim 100\times$ fainter fluxes with respect to coded-mask instruments. A wedding-cake strategy for the \nustar surveys was adopted: a shallow, wide-area survey of the COSMic Evolutionary Survey field \citep[COSMOS,][]{civano15}, a deep, pencil-beam survey of the Extended \chandra Deep Field-South \citep[ECDFS,][]{mullaney15}, and a serendipitous survey \citep{alexander13,lansbury17} were the first steps of a comprehensive survey program, which is now complemented by the observations of the Extended Groth Strip (EGS, Aird et al. in prep), \chandra Deep Field-North (CDFN, Del Moro et al. in prep) and UKIDSS - Ultra Deep Survey \citep[UDS,][]{masini18} fields.
\par Recently, \citet[M18 hereafter]{masini18} presented the results of a \nustar survey of the UKIDSS-UDS field. In addition to adopting the three most commonly used energy bands (F: $3-24$ keV, S: $3-8$ keV, H: $8-24$ keV), they explored the feasibility of source detection in three additional hard bands, splitting the commonly adopted hard band ($8-24$ keV) in two bands ($8-16$ keV and $16-24$ keV, H1 and H2 hereafter) and in a very-hard band ($35-55$ keV, VH hereafter), chosen as the energy interval where the \nustar background is mainly instrumental and well described by a fairly flat power-law \citep{wik14}. Splitting the H band into two sub-bands is important for two reasons. On one side, the background contribution is limited in the H1 band, allowing some sources to  be  more  significantly  detected  narrowing  the  band; on the other hand, selecting sources at $\sim16-24$ keV in  the  H2  band  helps  detecting  directly  those  AGNs contributing  the  most  to  the  peak  of  the  CXB\footnote{The H2 band energy range was chosen to preserve the original width of the \nustar $8-24$ keV band. Another possibility would have been to bridge the gap with the VH band and define the H2 band as H2$=16-35$ keV. It is not yet clear how the interplay between the larger band and the higher \nustar instrumental background above $\sim 25$ keV would impact the detection of hard sources in such a band.}. The VH band was chosen in order to exploit, for the first time in a deep extragalactic survey, the broad \nustar spectral coverage.
\par Few sources were detected in these bands, the majority of which were in the H1 band. Only one source was reliably detected in the H2 band, while no sources were detected in the VH band. These results were expected, given the intensity and shape of the \nustar background and the survey area and depth, thanks to a large set of simulations, run in order to maximize the reliability and completeness of source detection at the same time \citep[M18]{civano15}.
\par In this paper we exploit the homogeneity of the \nustar survey strategy adopted in its extragalactic surveys program, by performing source detection in the H1, H2 and VH bands also in the COSMOS and ECDFS fields, ultimately aggregating the results with those coming from the UDS field.
\par We assume a flat $\Lambda$CDM cosmology ($H_0 = 70$ km s$^{-1}$ Mpc$^{-1}$, $\Omega_{\rm M} = 0.3$, $\Omega_{\Lambda} = 0.7$) throughout the paper, which is organized as follows: the data sets and the sample are presented in Sections \S\ref{sec:datasets} and \S\ref{sec:sample}. We compute sensitivities in Section \S\ref{sec:sensitivities}, and use them to get the integral number counts in Section \S\ref{sec:lognlogs}. Section \S\ref{sec:cosmos12} presents an interesting source previously missed by \nustar and barely detected by \textit{Chandra} in the COSMOS field, and a source detected for the first time in the X-ray band in UKIDSS-UDS. We draw our conclusions in Section \S\ref{sec:conclusions}.

\begin{figure}
\plotone{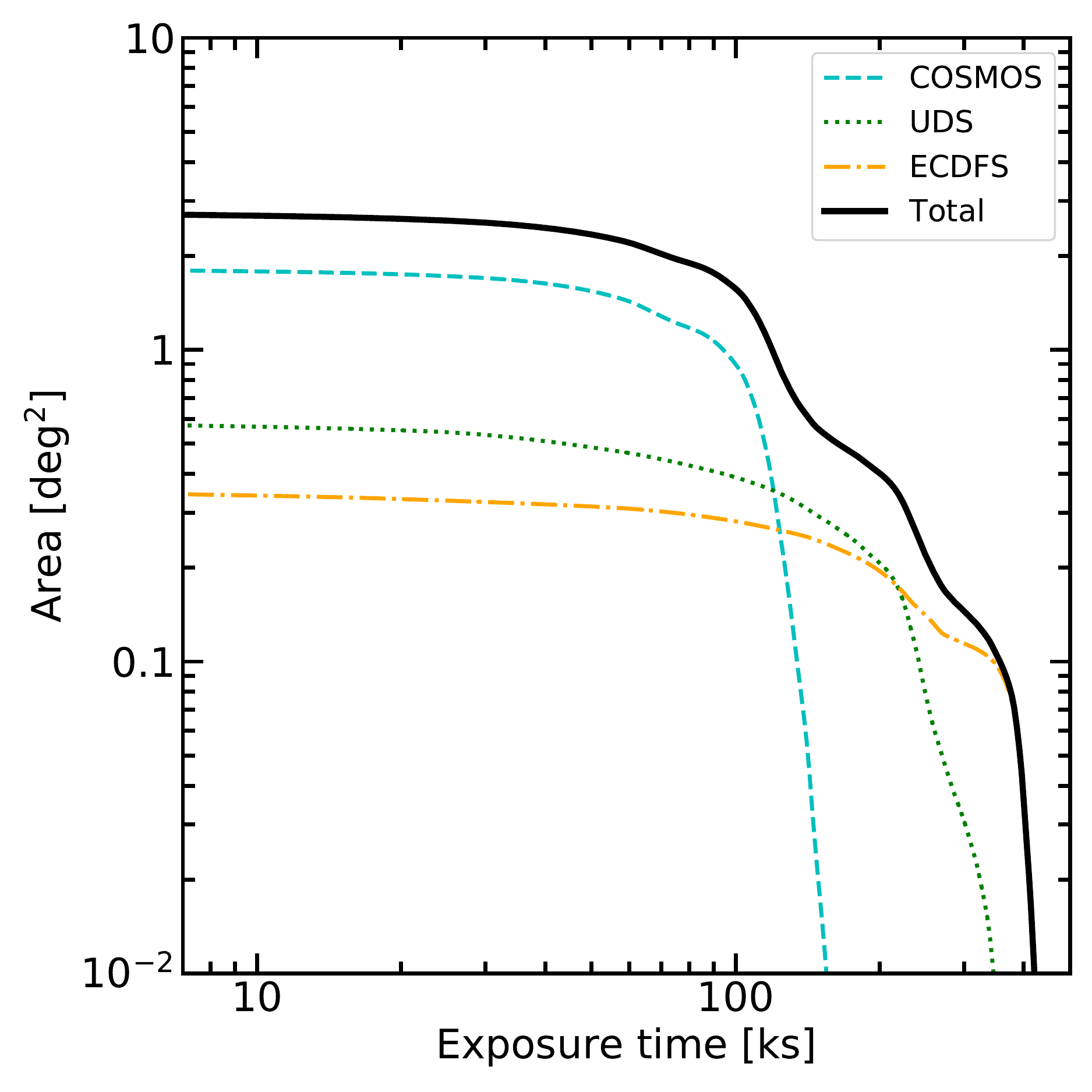}
\caption{Area as a function of FPMA+FPMB exposure for the three field considered in this work (COSMOS, dashed cyan;  UDS, dotted green; ECDFS, dot-dashed orange). The summed area curve is shown as the thick black line, for a total area of 2.74 deg$^2$. \label{fig:areavsexpo}}
\end{figure}

\begin{table}
\caption{Details on the single fields considered in this work.}             
\label{tab:det}      
\centering          
\begin{tabular}{l c c c c c c} 
\hline\hline       
\noalign{\vskip 0.5mm} 
 Field & R.A. & DEC.  & Area & $T_{\rm exp}$ & Ref. \\   
  & [deg] & [deg] & [deg$^2$] & [ks] & \\         
\noalign{\vskip 1mm}  
COSMOS & 150.2 & 2.2  & 1.81 & 155 & Civano+15\\ \noalign{\vskip 0.5mm}
UDS & 34.4 & -5.1 & 0.58 & 345  & Masini+18\\ \noalign{\vskip 0.5mm}
ECDFS & 53.1 & -27.8  & 0.35  & 420  & Mullaney+15\\ \noalign{\vskip 0.5mm}
\noalign{\vskip 1mm}    
\hline
\end{tabular}
\tablecomments{The area column refers to the total area, while the exposure time ($T_{\rm exp}$) column is the FPMA+FPMB exposure time at which the area drops to 0.01 deg$^2$. }
\end{table}

\section{Data sets} \label{sec:datasets}

We combine the \nustar surveys performed in three different fields (COSMOS, ECDFS, UDS; see Table \ref{tab:det}) in order to maximize the total survey area. In Figure~\ref{fig:areavsexpo}, the field areas (and the total one resulting from their sum) as a function of exposure are shown. Since creating the exposure maps in the H1 and H2 bands is time consuming, and the exposure maps in the \nustar H-band were already available, we adopted the H-band exposure maps for both the H1 and H2 bands\footnote{\citet{masini18} have shown that adopting the H-band exposure map for the H1 and H2 bands results in an underestimation of the exposure of at most 3\%, and overestimation of the exposure of at most 12\%, respectively.}, and created with the \texttt{nuexpomap} task the exposure maps in the VH band only, weighted at $E=44.35$ keV \citep[see][and M18, for further details]{civano15}.  The total area is 2.74 deg$^2$, with a half-area depth of 108 ks, with the two \nustar focal plane modules (FPMA and FPMB) summed together. 
\newline Details about the single surveys are available in the appropriate papers. In particular, we follow the same strategy as M18 in order to make the analysis as homogeneous as possible. Refer to the same paper for details on the data analysis, background maps production, runs of simulations, and detection strategy. The DET\_ML\footnote{This quantity is defined as DET\_ML$=-\ln(P)$, where $P$ is the Poissonian probability that the source is a background fluctuation. The higher the DET\_ML, the higher the significance of a source.} thresholds used in this work, corresponding to the 99\% and 97\% levels of reliability (defined through simulations as the ratio between sources matched with their counterpart, and detected sources, at a given DET\_ML), and the number of sources detected above a given threshold in each band are reported in Table \ref{tab:summary}. For example, in the simulations run in the COSMOS field, the $99\%$ of the detected sources with a DET\_ML $\geqslant 16.59$ was matched to their input counterparts (while the remaining 1\% was not matched, being made up of spurious sources).

\begin{table*}
\caption{Detection thresholds and numbers in each field and band.}             
\label{tab:summary}      
\centering          
\begin{tabular}{l c c c c c c | c  c c c c c c c c c c c} 
\hline\hline       
\noalign{\vskip 0.5mm} 
 Field &  \multicolumn{12}{c}{Reliability} \\
 &  \multicolumn{6}{c|}{99\% [Spurious fraction: 1\%]}  & \multicolumn{6}{c}{97\% [Spurious fraction: 3\%]} \\
 \hline
 &  \multicolumn{6}{c|}{Bands} & \multicolumn{6}{c}{Bands} \\   
  & \multicolumn{2}{c}{8-16 keV} & \multicolumn{2}{c}{16-24 keV} & \multicolumn{2}{c|}{35-55 keV} & \multicolumn{2}{c}{8-16 keV} & \multicolumn{2}{c}{16-24 keV} & \multicolumn{2}{c}{35-55 keV} \\
   &  Thr. & N & Thr. & N & Thr. & N  &  Thr. & N & Thr. & N & Thr. & N  \\  
\cline{2-13}            
COSMOS & 16.59 & 32 & 19.07 & 2 & 21.36 & 0 & 14.49 & 38 & 16.95 & 3 & 20.69 & 0 \\ 
UDS & 15.13 & 16 & 17.54 & 1 &  23.55 & 0 & 13.23 & 20 & 16.09 & 1 & 23.00 & 0\\ 
ECDFS & 16.81 & 11 & 19.09  & 0 & 28.54  & 0 & 14.22 & 14 & 16.86 & 0 & 28.38 & 0\\
\hline
Tot &  &59 & & 3 & & 0 & & 72 & & 4 & & 0 \\   
\hline
\end{tabular}
\tablecomments{The thresholds are expressed in terms of DET\_ML, defined as DET\_ML=$-\ln(P)$ where $P$ is the Poissonian probability that the source is a background fluctuation. The thresholds have been computed exploiting a large set of simulations, in order to keep under control the spurious fraction. See \citet{civano15} and M18 for further details.}
\end{table*}

\section{The aggregated sample}\label{sec:sample}

As can be seen from the last row of Table \ref{tab:summary}, regardless of the reliability threshold, the H2 and VH bands return very few or no detections at all. In particular, of four sources significantly detected at the 97\% level of reliability in the H2 band across the aggregated fields, no one is \textit{exclusively} detected in the H2 band. The majority of these sources (three) come from the COSMOS field, which is the widest of the set, while none comes from ECDFS, the deepest one. This is probably due to the rather high \nustar background in the H2 band, which requires sources to be very bright (and henceforth rare) in order to be robustly detected.  
\par Following M18, we focus on the largest sample, at the cost of a slightly lower reliability (spurious fraction of 3\%). The total sample of sources detected in at least one band is composed of 72 sources across the COSMOS (38 sources), UDS (20 sources) and ECDFS (14 sources) fields. 

\subsection{Matches with previous \nustar catalogs}\label{sec:matches}
Out of the 38 sources detected in the COSMOS field, 35 are matched with a counterpart from \citet{civano15} within $30''$. Such a matching radius has been adopted also in other \nustar Extragalactic Surveys \citep[M18]{civano15,mullaney15} and we refer the interested reader to those papers for more details. One of the three unmatched sources has its counterpart at a distance of $32''$, so slightly above our matching radius, and we consider this source as real, while the other two do not have a close counterpart. One of these two is significant above the 99\% of reliability threshold in the H1 band, and may have been missed by \citet{civano15} due to the background contribution in the H2 band (in which it is undetected) lowering the source significance in the aggregated $8-24$ keV band. We are going to focus on this source in \S\ref{sec:cosmos12}. The last one could be a spurious source, since its significance is just above the adopted threshold of 97\% of reliability. Indeed, from our spurious fraction of 3\% we expect an average of $\approx 1$ sources to be a false detection in the COSMOS field.
\par Nineteen of the 20 sources detected in the UKIDSS-UDS field are matched to the catalog of M18. The only source missing a counterpart has a DET\_ML just above the threshold, and may be a  spurious source as well. A discussion of the possible counterparts of this source is presented in \S\ref{subsec:uds10}. 
All the 14 sources detected in the ECDFS are matched to the catalog of \citet{mullaney15}.
\par Given these results, one may think that the sources detected in the H1 band are the same ones detected in the broader H band, giving no reason to prefer the H band over the H1 band. However, some differences appear at a closer look. Indeed, if the narrower band suffers from less background, it also loses sensitivity over the broader one. We stick to the 99\% reliability threshold to have an easier and more homogeneous comparison with the UDS, COSMOS and ECDFS catalogs. As can be seen from Table \ref{tab:H1}, there are 32 sources detected in the COSMOS field in the H band, and 32 in the H1 band as well. However, only 26 of these are in common. In other words, employing the H1 band six more sources are detected, but six  sources are lost at the same time. Similarly, in UDS there are 15 sources detected in the H band, and 16 in the H1 band. In this case, 13 sources are in common: preferring the H1 band, three sources are gained, but two sources are lost. In ECDFS, there are 19 sources in the H band, and 11 in the H1 band. All the 11 H1 sources are detected in the H band, implying a loss of 8 sources.
\par We conclude that there is no clear benefit in employing the H1 band over the H band in a deep \nustar survey, at least performing detections at the 99\% reliability level. However, as we shall see in the following Sections, there are few cases in which the H1 band is extremely helpful in selecting interesting sources with a particular spectral shape.

\subsection{Other details of the sample} \label{sec:details}
Our final sample is made of 72 sources. Every source in this sample is detected above the threshold of 97\% reliability in the H1 band, four of these sources are also significantly detected in the H2 band, while no sources are detected above the threshold in the VH band. Two sources (id158 and id218 in the catalog of \citealp{civano15}) have a sub-threshold counterpart in the VH band, and are both detected in the COSMOS field. This means that a detection in the VH band has been found for these two sources, albeit their DET\_ML values in the VH band are not sufficient to claim a detection at the 97\% reliability level. In particular, the first one (DET\_ML$_{\rm VH}$ $\sim 11.66$) is associated with a quasar at $z=1.509$ (spectroscopic), is detected above the threshold in the H1 band but undetected in the H2 band, while the second one (DET\_ML$_{\rm VH}$ $\sim 8.1$, associated with a quasar at a spectroscopic redshift of $z=0.345$) is robustly detected in both the hard sub-bands H1 and H2, giving further support to the hypothesis that the weak emission in the VH band could be associated with such a bright source.

\begin{figure}
\plotone{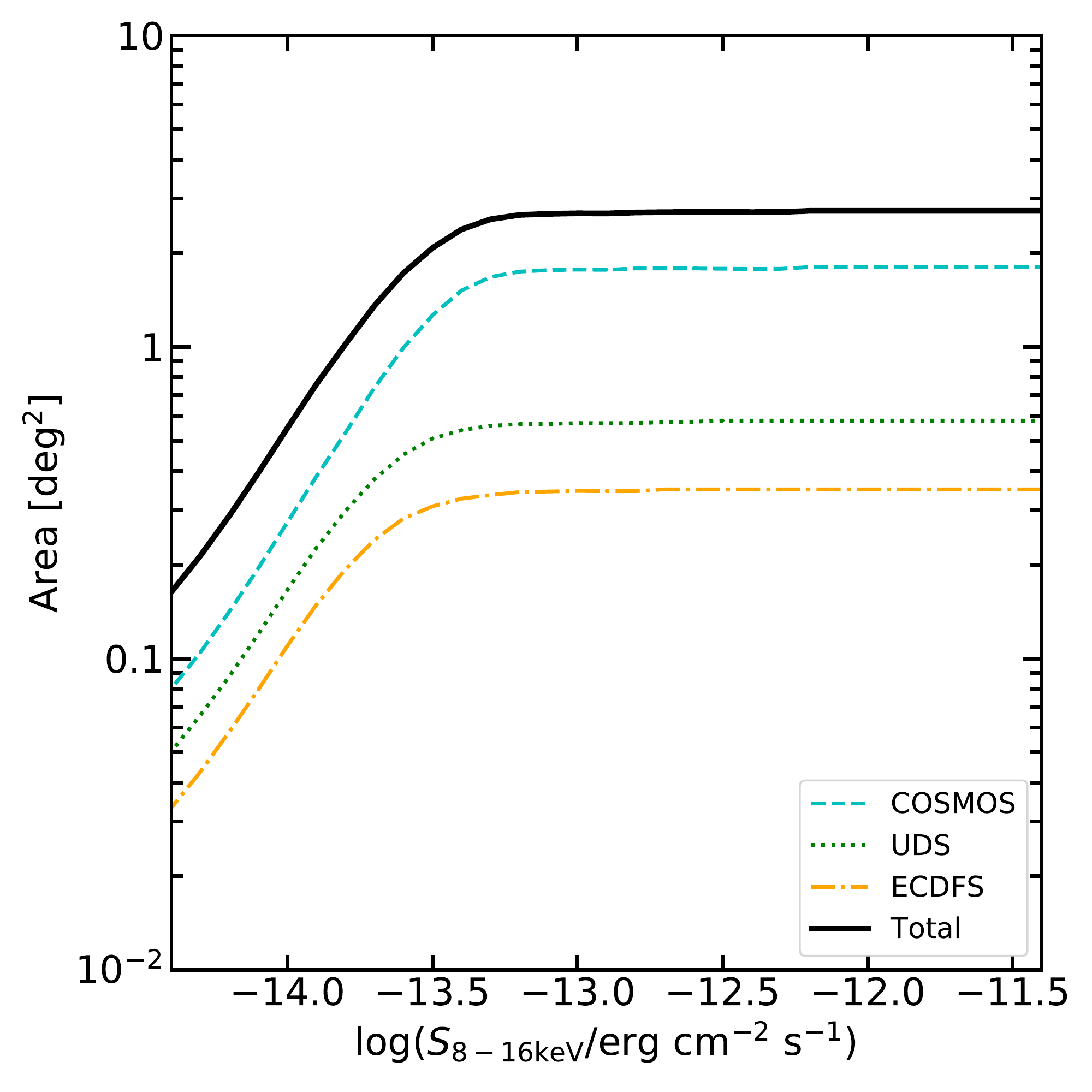}
\plotone{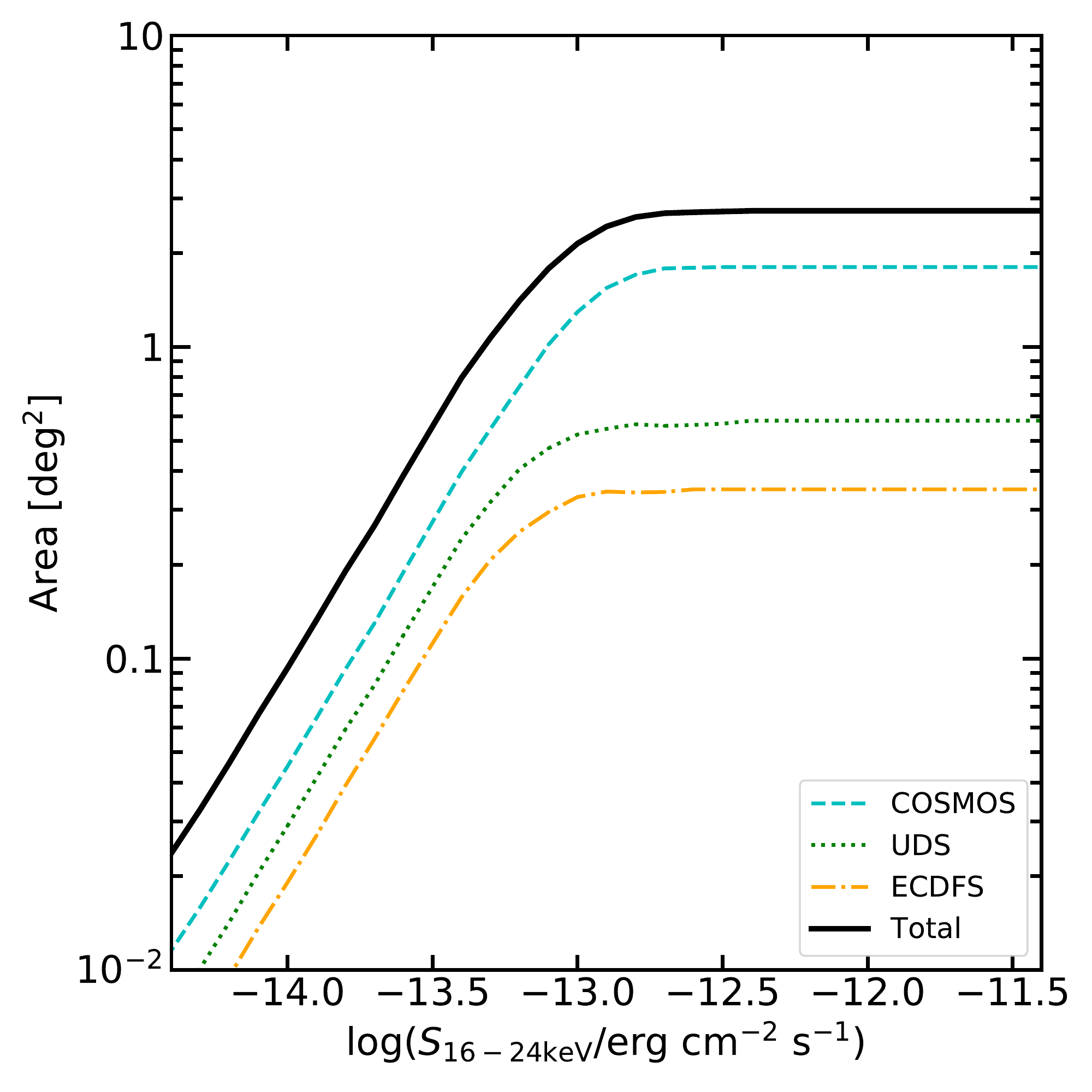}
\plotone{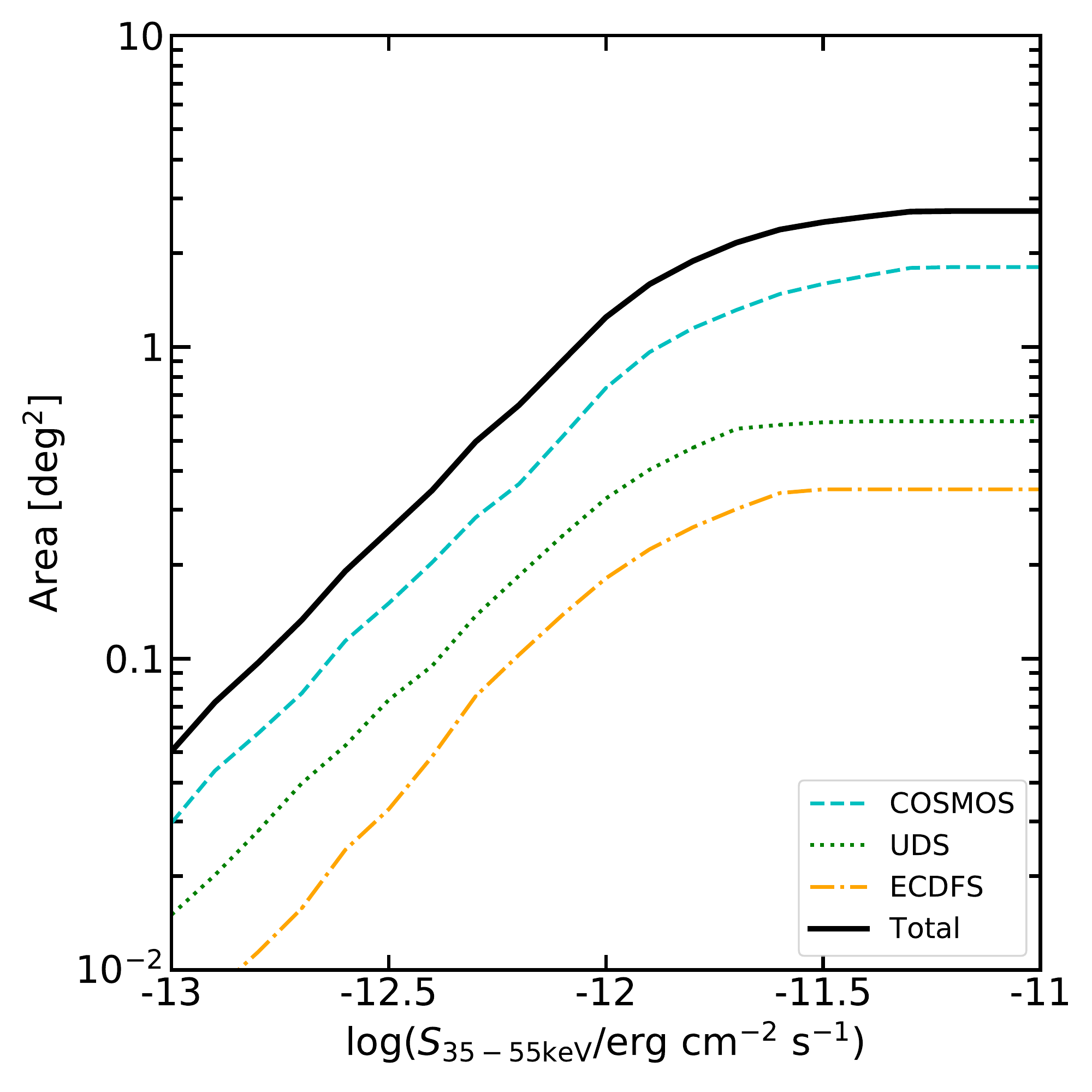}
\caption{From top to bottom: sensitivity for the H1, H2, and H3 band, respectively. In each panel, the COSMOS curve is shown by the cyan dashed line, the UDS one in dotted green, and the ECDFS one in dot-dashed orange. The total curve is the black thick line. \label{fig:sens}}
\end{figure}

\section{Sensitivities}\label{sec:sensitivities}

In order to adequately take into account the survey sensitivity we compute, for each field and each band, the completeness at the 97\% reliability threshold. Further details on the definitions of reliability and completeness for our surveys are provided in \citet{civano15} and M18. We briefly recall here that they both exploit the large suite of simulations run for each energy band and each field. The reliability sets the threshold above which a source is considered to be real, keeping under control the spurious fraction. The completeness is computed as the ratio of the number of sources detected above a defined reliability threshold and matched to their counterparts, and the number of sources injected in a simulation, as a function of their input flux. In other words, at high fluxes the completeness curve is unity, because bright sources are easily detected and matched to their input counterparts; at fainter fluxes, more and more sources are missed (and spurious sources are more easily detected), lowering the survey completeness. Rescaling this curve for the total area, a sensitivity curve is obtained, which naturally encompasses the Poissonian fluctuations of the background at low fluxes \citep[e.g.,][]{georgakakis08}. 
\par This procedure has been adopted in an identical way for the three fields and for each band\footnote{Due to the rarity of bright sources in the VH band in the smallest fields (UDS and ECDFS), a total of 1000 simulations were run. We ran 400 simulations for the other bands, consistently with M18.}, and the total area curve is obtained by summing the sensitivity curves of the three fields. The sensitivities for the H1, H2 and VH bands are shown in Figure \ref{fig:sens}.

\begin{table}
\caption{Balance between H and H1 bands at the 99\% reliability level in the three fields considered. The Gained/Lost column refers to adopting the H1 band over the H one.}             
\label{tab:H1}      
\centering          
\begin{tabular}{l c c c c c c} 
\hline\hline       
\noalign{\vskip 0.5mm} 
 Field & H & H1  & In common & Gained/Lost \\       
\noalign{\vskip 1mm}  
COSMOS & 32 & 32  & 26 & +6/-6 \\ \noalign{\vskip 0.5mm}
UDS & 15 & 16 & 13 & +3/-2 \\ \noalign{\vskip 0.5mm}
ECDFS & 19 & 11  & 11  & 0/-8 \\ \noalign{\vskip 0.5mm}
\noalign{\vskip 1mm}    
\hline
\end{tabular}
\end{table}

\begin{figure}
\epsscale{0.85}
\plotone{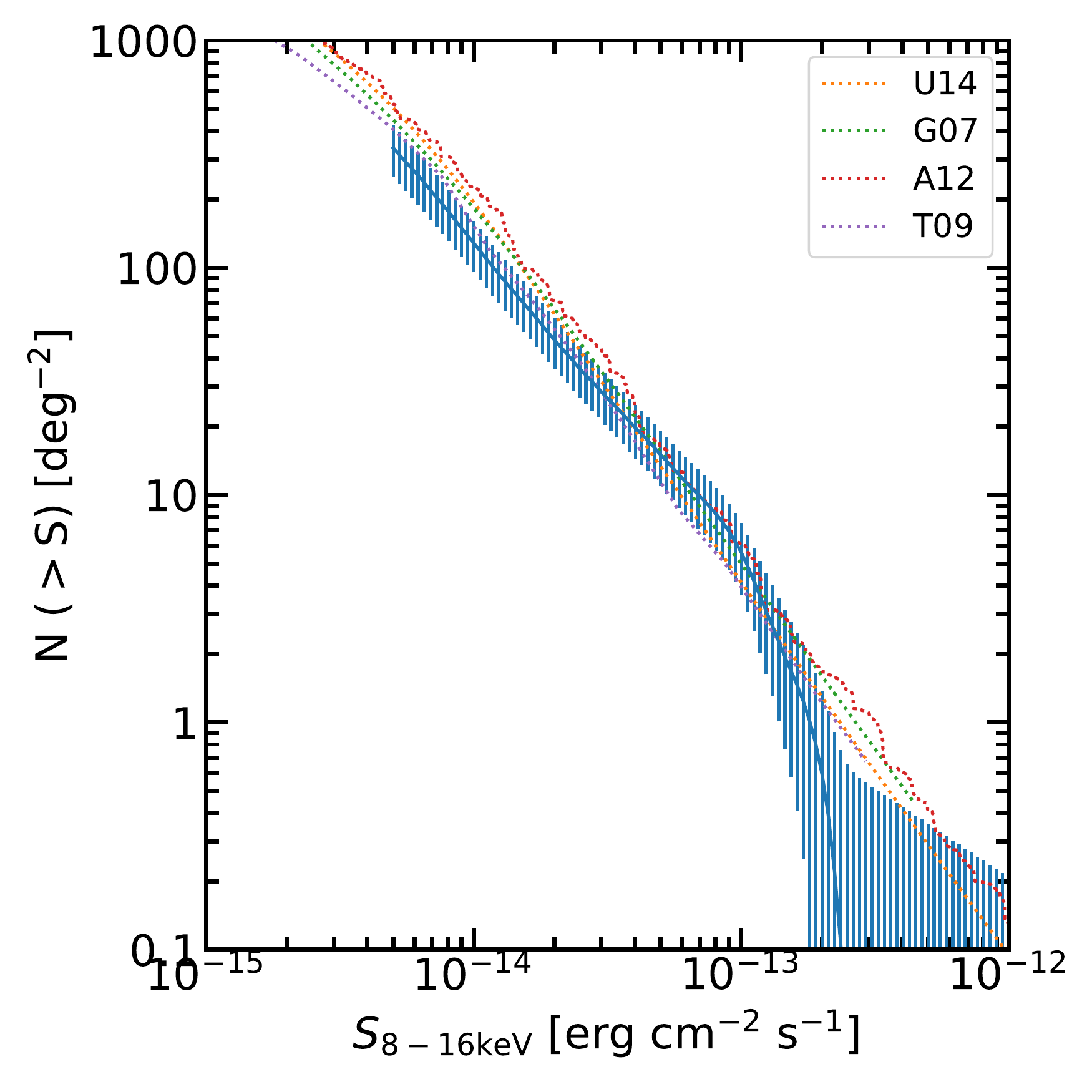}
\plotone{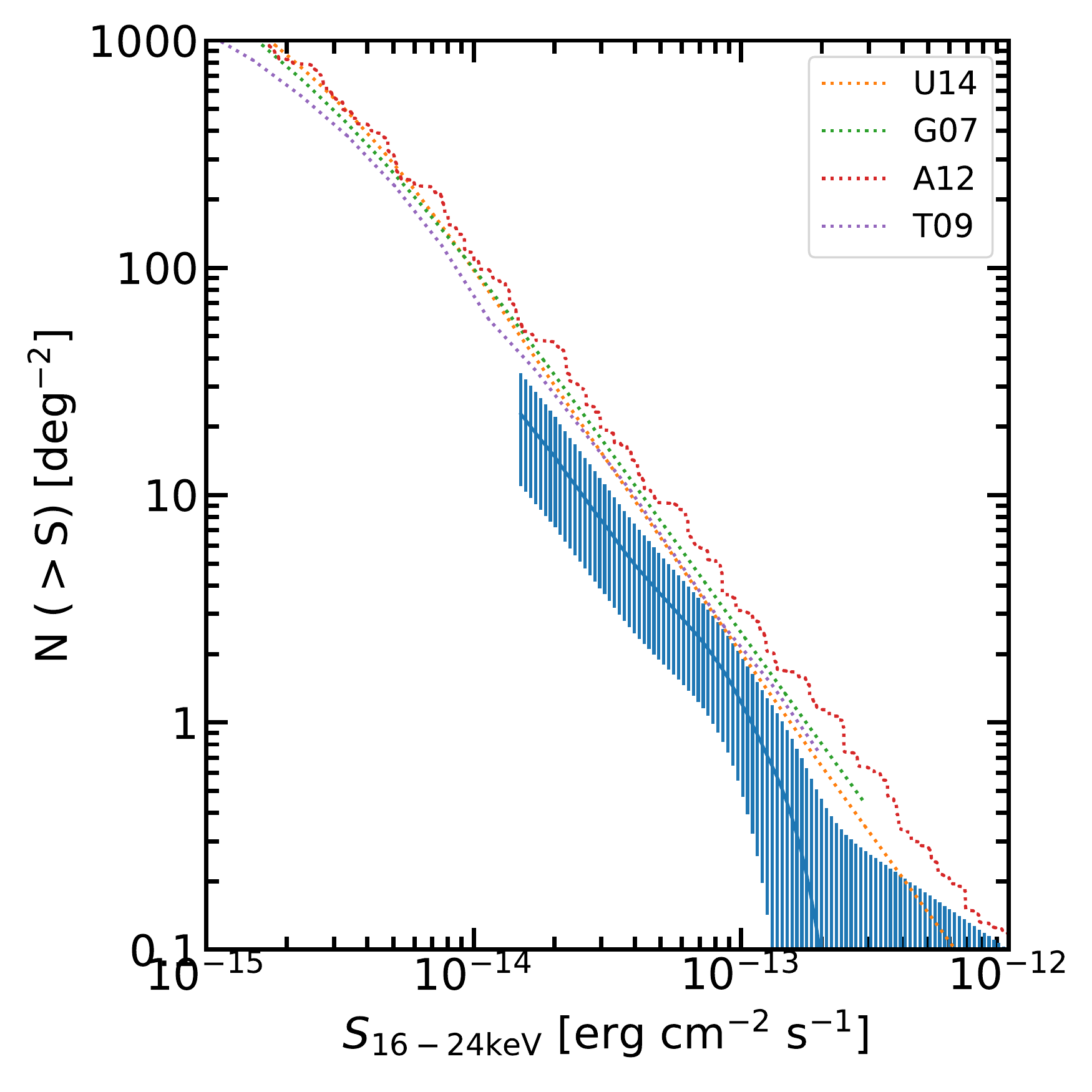}
\plotone{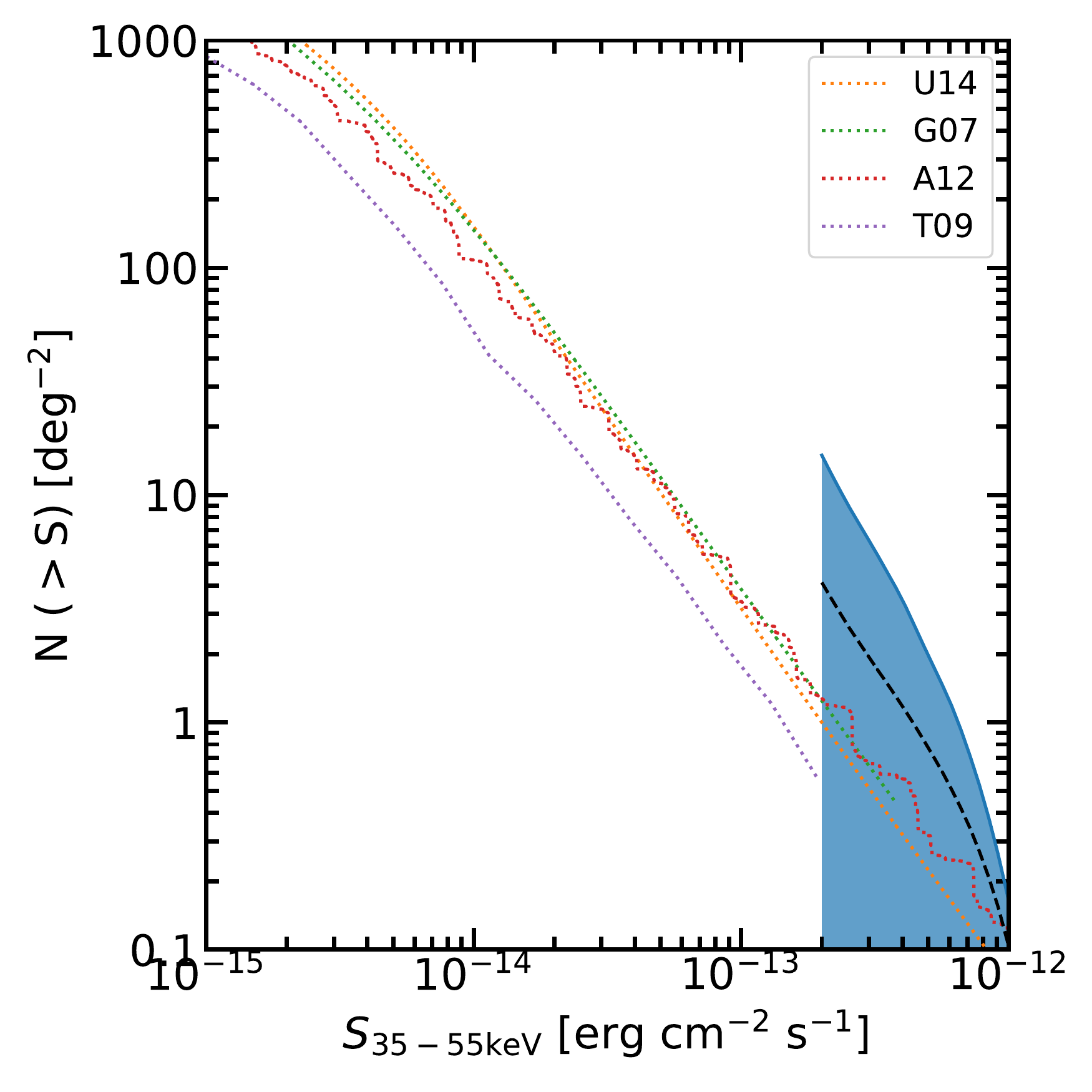}
\caption{From top to bottom: $\log N-\log S$ for the H1, H2 and upper limits for the VH band, compared with some population synthesis models (in dotted lines: orange, \citealt{ueda14}; green, \citealt{gilli07}; red, \citealt{akylas12}; violet, \citealt{treister09}). The uncertainties are Poissonian, and take into account a rough constant 13\% of cosmic variance in the H1 band. In the bottom panel, the dashed black line denotes where the upper limit would lie adopting lower reliability thresholds, appropriate for the two tentative detections. \label{fig:lognlogs}}
\end{figure}

\section{LogN-LogS} \label{sec:lognlogs}

For every source and in each band, we extract total counts, background counts, and average exposures from a circular aperture of $20''$ from the data mosaics, background mosaics and exposure maps, respectively. Detections and non-detections are treated in the same way of \citet{civano15} and M18. \par Following \citet{harrison16}, we adopt a Bayesian method in calculating the number counts of our sources \citep{georgakakis08}. Briefly, instead of assuming a fixed flux (coming, e.g., from aperture photometry) with no uncertainty, we compute the probability density function for each source of having a flux $f$ inside a defined range of fluxes. The minimum flux is a factor of three lower than the flux limit at the 50\% of completeness in each band as reported by M18, while the maximum flux is $10^{-12}$ \fluxcgs for the H1 and H2 bands, and $10^{-11}$ \fluxcgs for the VH band. For each source, the expected number of total counts in order to have a flux $f$ is 
\begin{equation} 
T=ft_{\rm exp}C\eta + B,
\end{equation} 
where $t_{\rm exp}$ is the exposure time of the source, $C$ is the conversion factor between fluxes and count rates (assuming an average photon index $\Gamma=1.8$; see, e.g., \citealp{burlon11}), $\eta$ is the factor to take into account the encircled energy fraction of the PSF (for \nustar and the aperture of 20$''$ adopted to extract counts, this results in $\eta=0.32$; see \citealt{civano15}) and $B$ are the background counts. The probability of having $T$ total counts given the observed total counts $N$ is then \citep[see Eq. 6 in][]{georgakakis08}: 
\begin{equation} 
P(f,N)=\frac{T^N e^{-T}}{N!}f^{\beta}.
\end{equation} 
Here $\beta$ is the slope of the differential number counts, and we assume $\beta=-2.81$ following \citet{harrison16}. The exact choice of $\beta$ has a negligible impact on the slope of our integral number counts: varying $\beta$ by 40\% results in a 3\% variation of the integral number counts slope. \par
We normalize such density functions in order to have a unitary contribution of a source split on the whole flux range. Summing the single probability density functions for each source and dividing by the sensitivity curve gives the number counts. The integral number counts for sources detected by \nustar in the H1, H2 and VH bands are shown in Figure \ref{fig:lognlogs}, along with predictions from some population synthesis models. 
\par In the top panel of Figure \ref{fig:lognlogs}, the cumulative number counts of our H1-detected sources is presented. Our Poisson uncertainties, which include a constant $\sim13\%$ 
of cosmic variance for the H1 band, estimated following \citet{hickoxmarkevitch06}\footnote{We also validated our method exploiting the XMM-Newton Stripe 82 catalog \citep{lamassa16}. We measured the variance of sources extracted from randomly chosen circular areas of 2.7 deg$^2$, applying appropriate flux cuts in the $0.5-10$ keV band of $>3\times10^{-14}$ erg cm$^{-2}$ s$^{-1}$, 
broadly comparable with the \nustar limiting $8-16$ keV 
flux of our detected sources, respectively.}, suggest that we compare broadly well with models. On the other hand, a comparison of our number counts in the H2 band with models (middle panel of Figure \ref{fig:lognlogs}) shows that our data lie a factor of $\sim 2$ below the models' predictions. While this tension is mild, the systematic overestimation from models could suggest that \nustar is not detecting all the H2-band sources it should, or that models predict too many sources to be detected in the H2 band. The very limited number of sources (four, above the threshold of 97\% of reliability) prevents a more detailed discussion.
Finally, assuming the signal in the VH band to be real for the two COSMOS sources, we can place an upper limit on the number counts of sources in the VH band. The upper limit shown in the bottom panel of Figure \ref{fig:lognlogs} is to be considered overestimated, since the two tentative detections have a significance lower than the threshold for which the sensitivity curve has been computed. As noted in Section \ref{sec:details}, the two tentative detections have DET\_ML values of $\sim 8.1$ and $\sim 11.6$. If we lower the reliability thresholds of the three fields, so that sources with DET\_ML $\gtrsim 8.0$ are considered robust detections, the area seen at a given flux increases, at the cost of a lower reliability ($\sim 20-30\%$ across the fields). The larger area surveyed at a given flux reflects in a lower value of the upper limit on the integral number counts, indicated in the bottom panel of Figure \ref{fig:lognlogs} by the dashed black line.

\begin{figure*}
\centering
\includegraphics[width=\textwidth]{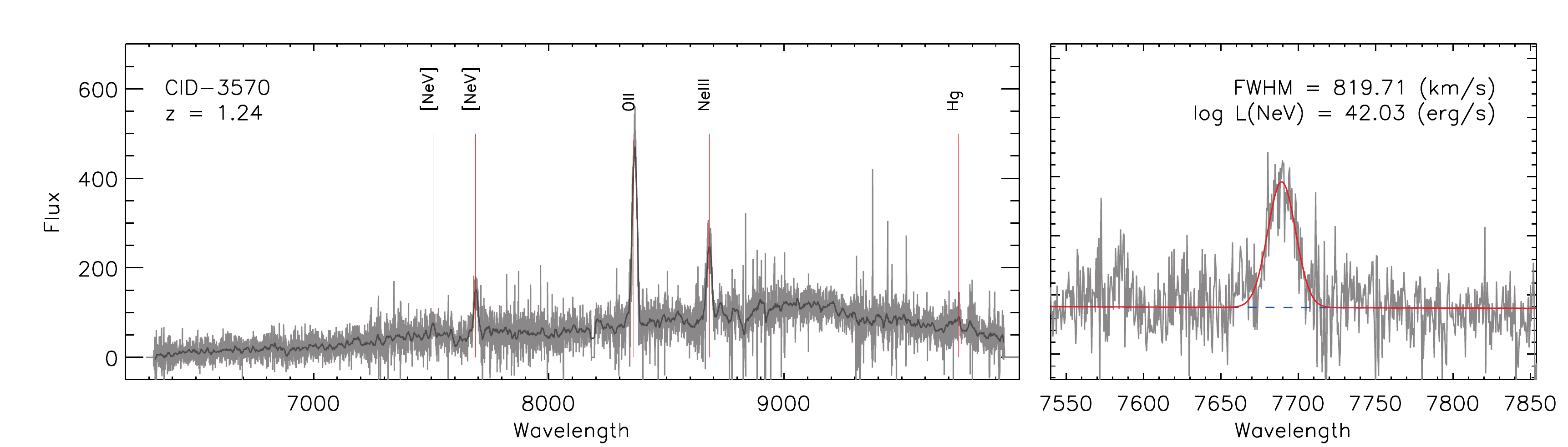}
\includegraphics[width=0.9\textwidth]{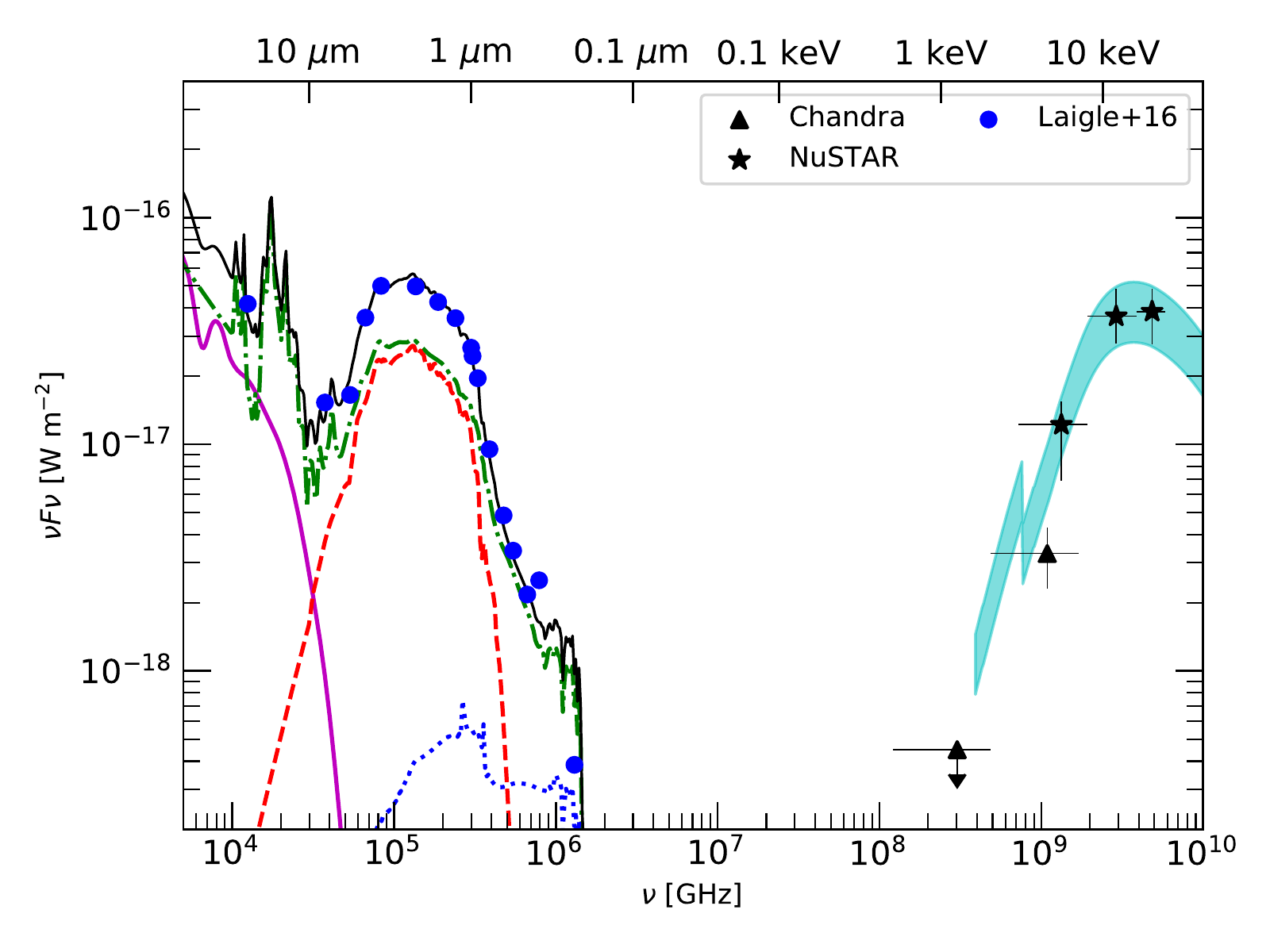}
\caption{\textbf{Top.} DEIMOS spectrum of  \cid \citep{hasinger18}, with the right panel zooming on the NeV emission line.  \textbf{Bottom.} Broadband SED for \cid, the interesting source at $z=1.244$ in the COSMOS field significantly detected in the H1 band by \nustar. The optical-IR part of the SED is fitted with an AGN \citep[solid violet,][]{assef10}, passive  \citep[dashed red,][]{assef10},  late type  \citep[dot-dashed green,][]{kirkpatrick15},  and star-forming  \citep[dotted blue,][]{assef10} templates. Data points (blue circles) are from the catalog of \citet{laigle16}, and references therein. In the X-ray part of the SED, the \textit{Chandra} (triangles) soft ($0.5-2$ keV) band flux is an upper limit. The \nustar fluxes derived from the spectral analysis are labeled with stars. Uncertainties on $\nu F \nu$ are at 1$\sigma$ confidence level, and the cyan region is the \texttt{pexrav} \citep{magdziarzzdziarski95} model ($\pm 1\sigma$) adopted to derive the \nustar fluxes. \label{fig:cosmos12}}
\end{figure*}

\section{A buried AGN in the COSMOS field}\label{sec:cosmos12}
As already discussed in \S\ref{sec:matches}, one of the two sources in the COSMOS field which is not matched with the catalog of \citet{civano15} is above the threshold of 99\% of reliability in the H1 band, and as a consequence would have been detected also in the most conservative sample of 59 sources, where the spurious fraction is $\sim 1\%$ (see Table \ref{tab:summary}). While missing a previously-detected \nustar counterpart\footnote{We note that a source at 13$"$ from our \nustar position is detected by \citet{civano15} in the full and hard \nustar bands. However, being below the 99$\%$ reliability threshold, it was not included in the final catalog.}, a faint \textit{Chandra} source is found at $\sim 7''$ distance in the \textit{Chandra} COSMOS Legacy Survey \citep[\cid in the catalog of][]{civano16}. \textit{Chandra} detected the source with $\sim16$ full ($0.5-7$ keV) band net counts, of which $\sim13$ are in the hard ($2-7$ keV) band. As a consequence, this source has an hardness ratio $\gtrsim0.68$ \citep{civano16}. In the optical band, \cid is associated with a galaxy at redshift $z=1.244$ \citep{kartaltepe15, marchesi16}. We thus expect the \nustar detection to be real, and associated with a highly obscured AGN with a hard spectrum barely detected by \textit{Chandra} and previously missed by \nustar. The \nustar spectrum of this source is indeed very hard, and a fit with a power law returns a flat photon index ($\Gamma = 0.56^{+0.96}_{-1.03}$), and a default MYTorus\footnote{The parameter describing the column density $N_{\rm H}$ in this model is in the range $10^{22}-10^{25}$ cm$^{-2}$; as such, if a measurement of $N_{\rm H}$ hits one of the caps, a letter $+u$ or $-l$ in the uncertainty is adopted for upper and lower caps, respectively.} model \citep{murphy09}, assuming a photon index $\Gamma=1.8$, gives a column density $N_{\rm H} = 2.3^{+u}_{-1.7} \times 10^{24}$ cm$^{-2}$ at the 90\% confidence limit. We notice that a pure reflection model \citep[\texttt{pexrav},][]{magdziarzzdziarski95} also gives a good fit, statistically indistinguishable from the MYTorus model adopted to derive an estimate on the column density. In particular, the photon index is fixed to $\Gamma=1.8$, and the reflection parameter is fixed to a negative value in order to have a pure reflection component. All the other parameters are left frozen to their standard values. With this configuration, the only free parameter of the \texttt{pexrav} model is the normalization at 1 keV, which is $N = 2.6^{+1.4}_{-1.2} \times 10^{-5}$ photons/keV/cm$^2$/s. Leaving the photon index free to vary does not improve the fit, and results in $\Gamma=1.4 \pm 1.0$. \newline
The DEIMOS \citep{faber03} spectrum of \cid \citep{hasinger18} is shown in the top panels of Figure \ref{fig:cosmos12}, where a clear [NeV]3426 emission line can be seen, leaving no doubts on the presence of an AGN. The X-ray to [NeV] luminosity ratio ($L_{\rm X}/L_{\rm [NeV]} \sim 17$), and the column density measured from the spectral analysis place \cid in very good agreement with the expected trend of the X/NeV ratio identified by \citet{gilli10} using a sample of local objects.
 \par In order to have a better view of the properties of \cid, we collected optical and near-IR data between 0.44 $\mu m$ and 24 $\mu m$, from the catalog of \citet{laigle16}. In particular, we use data from GALEX (NUV), CFHT ($u$), the Hyper Suprime Cam Subaru survey (B,V,$r$,$i^+$,$z^{++}$,Y), UltraVISTA (Y,J,H,K), and \textit{Spitzer} (IRAC and MIPS 24). We also add information in the X-ray band coming from \textit{Chandra} ($0.5-2$ keV and $2-7$ keV; see \citealp{civano16}, \citealp{marchesi16} and \citealp{marchesi16c}) and \nustar (S, H1, and H2 bands). \chandra fluxes are rescaled based on the count rate from \citet{civano16}, assuming the best-fitting \texttt{pexrav} model and simulating a fake \chandra spectrum with the appropriate response and ancillary files \citep{marchesi16c}. \nustar fluxes are computed through spectral analysis and adopting the reflection (\texttt{pexrav}) model, and would be the same within the uncertainties adopting the MYTorus model. We show the Spectral Energy Distribution (SED) in the bottom panel of Figure~\ref{fig:cosmos12} where the optical-IR part of the SED is fitted with a combination of four templates (Carroll et al., in prep): one for the AGN \citep{assef10}, and three for the host galaxy \citep{assef10,kirkpatrick15}. Through a $\chi^2$ minimization, the software returns the best fit among all the possible combinations of the four templates. The X-ray part of the SED, on the other hand, is fitted with the \texttt{pexrav} model described before. 
 \par  As can be seen from Figure \ref{fig:cosmos12}, the AGN component is required to better fit the MIPS 24~$\mu m$ and IRAC 8~$\mu m$ points only, and as a consequence is extremely reddened ($E(B-V)\sim 22.3)$. \par On the other hand, the inclusion of a cold dust component \citep{mullaney11} in the fitting \citep[e.g., ][]{suh17}, due to the source being detected in the FIR band at 250 $\mu m$, may reduce the AGN contribution at 24$\mu m$, and its relative reddening ($E(B-V) \sim 0.3$).
\par Such different results, dependent on the models adopted, point toward a general lack of evidence of the AGN component in the UV/optical/IR SED of \cid, giving on the other hand more importance to the X-ray detection. Indeed, regardless of the exact details of the SED fitting, the main point of this analysis is that the \nustar detection in the H1 band is associated with a real, highly obscured AGN. This motivates a deeper look to the unassociated source found in the UDS field as well.

\subsection{Another buried AGN in UKIDSS-UDS?}\label{subsec:uds10}
Motivated by the successful confirmation of a Compton-thick AGN for the unassociated source in the COSMOS field \cid, we focused on the source not matched with the catalog of M18 in the UDS field. \par Within 30$''$ from the H1-band position, we found one faint \textit{Chandra} source (Kocevski et al. 2018, in press) at a distance of $\sim12''$, associated with a galaxy in a high-redshift cluster (namely CVB13-22, at spectroscopic redshift $z=1.4548$; see \citealt{vanbreukelen07}). \par
\begin{figure*}
\centering
\includegraphics[width=0.95\textwidth]{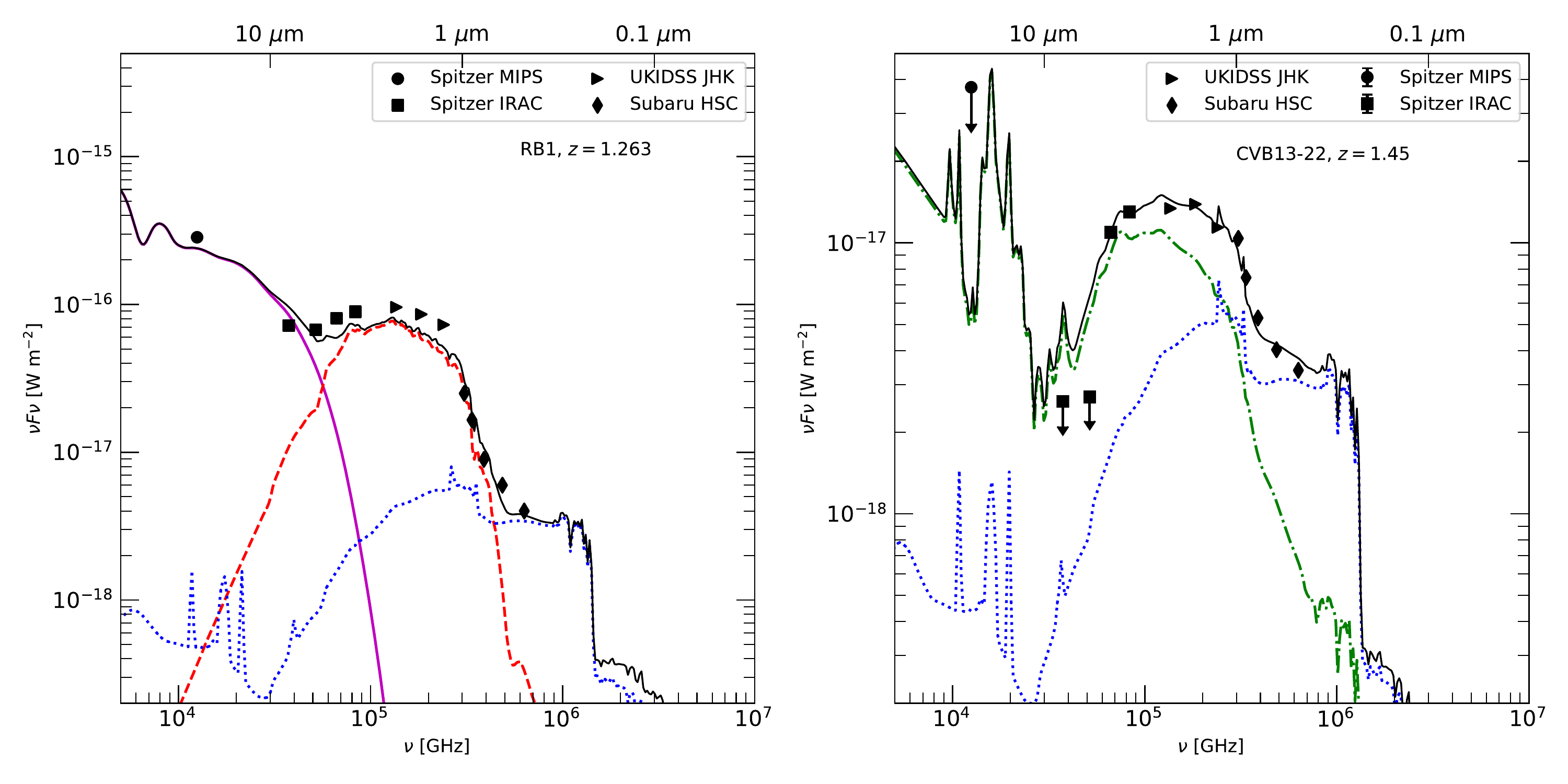}
\caption{Optical/IR SEDs of RB1 (left) and CVB13-22 (right), the two possible counterparts to the \nustar detection in the UDS field. The circle datapoint is \textit{Spitzer} MIPS, the squares refer to \textit{Spitzer} IRAC, the rightward triangles to UKIDSS in the JHK bands, and the diamonds to Subaru HSC in the $g,r,i,z,Y$ bands.  The optical-IR part of the SED is fitted with an AGN (solid violet), passive (dashed red), late-type (dot-dashed green) and star-forming (dotted blue) templates. Upper limits are labeled as downward arrows, and have been estimated considering the minimum flux of the sources detected by one instrument in the surroundings of the target. While the AGN component in the SED in RB1 is heavily reddened ($E(B-V)\sim 6.2$), giving support to the classification of obscured (possibly CT) AGN, the same component in CVB13-22 is not required by the fit. \label{fig:uds10}}
\end{figure*}

 On the other hand, at approximately the same distance ($\sim12''$), we find a second possible counterpart to the \nustar detection. Indeed, \citet{simpson12} classify the source RB1 in \citet{vanbreukelen07} as a Narrow Line AGN (NLAGN) at spectroscopic redshift $z=1.263$. The classification comes from \citet{chuter11}, and is due to the detection of strong, high-ionization UV emission lines like C IV and [Ne III]. The steep NIR-MIR slope in the SED of RB1, contrary to the non-detection at the same wavelength of CVB13-22 (see Figure \ref{fig:uds10}) argues in favor of this AGN being the correct counterpart, although we cannot rule out the possibility that also the CVB13-22 source in the high-redshift cluster could be at least partially contributing to the \nustar flux. If RB1 is the correct or dominant counterpart, the \nustar detection would be the first X-ray detection of this AGN. In either case, we consider also this detection to be real, giving further support to the use of the H1 band as a tool to unveil candidate buried AGNs previously missed by X-ray surveys.

\section{Future prospects for the H1 and H2 bands}
As already discussed (\S\ref{sec:matches}), the H1 band is suitable to find interesting, rare and obscured AGN, but misses some of the H-band selected sources, due to the loss of sensitivity narrowing the band. Of a total of 72 H1-band sources detected, three were not matched to any \nustar source previously detected in the F, S or H bands. Two out of three turned out to have an optical counterpart, showing signatures of buried AGNs at $z > 1$. It is very likely that \cid in COSMOS (\S\ref{sec:cosmos12}) would have been robustly detected in the usual F and H bands by \citet{civano15} relaxing the reliability threshold. On the other hand, RB1 in the UDS field (\S\ref{subsec:uds10}) was missed by M18 even employing the less-conservative thresholds for the F, S and H bands. Given the large overlap with the H-band selected sources, we suggest to employ the H1 band for follow-up observations of IR-selected potentially buried AGN.
\par Regarding the four sources detected in the H2-band, they have broadly the same H2-band flux ($F_{\rm H2}\sim1-2\times10^{-13}$~\fluxcgs). At this flux, the total area is $\sim2.45$ deg$^2$, with an average exposure time $\sim 40$ ks (FPMA+FPMB). This implies that we found $\sim1.6$ H2-band sources/deg$^2$, or equivalently one H2-band source for each $\sim 0.63$ deg$^2$. This number is remarkably consistent with the one found in UDS ($\sim 0.6$~deg$^2$), three in COSMOS ($\sim 1.8$~deg$^2$), and none in ECDFS ($\sim 0.4$~deg$^2$). These numbers thus imply that \nustar is able to detect in the H2 band one source every $\sim23$ pointings of 20 ks, if they are not overlapping. If overlapping with the half-FOV strategy commonly adopted in \nustar surveys, 50\% more pointings are needed, requiring $\sim34$ half-FOV shifted pointings to detect one source. Indeed, COSMOS is made up of 121 partially overlapping pointings, and should contain $\sim 3.6$ sources; UDS is made up of 35 half-FOV shifted pointings, just enough to detect one source. This is not happening for the ECDFS survey, which has two totally overlapping passes on a half-FOV shifted $4\times4$ square.
The ever-growing \nustar Seredipitous Survey \citep{alexander13,lansbury17}, with its $\sim12.5$ deg$^2$ coverage at the minimum flux of our sources, should contain $\sim 20$ sources detectable in the H2 band by \nustar in $\sim20$ ks.

\section{Conclusions}\label{sec:conclusions}
In this paper, we presented the aggregated results coming from three different \nustar survey fields in three hard bands (H1: $8-16$ keV, H2: $16-24$ keV, VH: $35-55$ keV), covering a total area of $2.7$ deg$^2$. The main results can be summarized as follows:
\begin{itemize}
\item{Following the same strategy delineated in M18, we detected 72 sources above the 97\% level of reliability in at least one band across three fields. All the 72 sources are robustly detected in the $8-16$~keV band, while four of them are also detected in the $16-24$~keV band and two are perhaps detected (albeit under-threshold) in the $35-55$~keV band.  The expected spurious fraction of the aggregated sample is 3\%.}
\item{We computed the number counts for our sources. We took into account the Eddington bias in our sample, computing for each source a probability density function over a range of fluxes, i.e. allowing each source to contribute to each flux bin following its probability density function. A comparison with AGN population-synthesis models shows broad consistency with our results, mainly in the H1 band. A tension of a factor of $\sim 2$ in the H2 band is seen between the data and the models. Upper limits are provided in the VH  band, assuming the two sub-threshold detections to be real.}
\item{Narrowing the \nustar hard ($8-24$ keV) band and employing the H1 band can help selecting interesting and obscured sources previously missed due to the high background rising above $\sim15$ keV. We found at least one such example of a buried, likely Compton-thick AGN in the COSMOS field at $z\sim1.25$, together with an elusive AGN in the UDS field. Despite the presence of the [NeV] line and the \nustar spectral analysis, the SED of the source in COSMOS does not show strong evidence for the presence of and AGN. On the other hand, the most likely counterpart of the UDS detection shows a significant AGN component in its optical-IR SED.}
\item{Based on the results obtained in the H2 band, we computed that one H2-band source is detected every $\sim0.6$ deg$^2$, or equivalently every $\sim23$ non overlapping \nustar  pointings. We predict that the \nustar Serendipitous Survey should contain $\sim$ 20 H2-band sources robustly detectable by \nustar in a minimum time of 20 ks.}
\end{itemize}

\acknowledgments
We thank the anonymous referee, whose comments helped improving the quality and clarity of the paper. \newline
This work was supported under NASA Contract NNG08FD60C, and made use of data from the \nustar mission, a project led by the California Institute of Technology, managed by the Jet Propulsion Laboratory, and funded by the National Aeronautics and Space Administration. We thank the \nustar Operations, Software, and Calibration teams for support with the execution and analysis of these observations. This research made use of the \nustar Data Analysis Software (NuSTARDAS) jointly developed by the ASI Science Data Center (ASDC, Italy) and the California Institute of Technology (USA). This research has also made use of data obtained from the \chandra\ Data Archive and the \chandra\ Source Catalog, and software provided by the \chandra\ X-ray Center (CXC).
Figure \ref{fig:cosmos12} is partially based on data products from observations made with ESO Telescopes at the La Silla Paranal Observatory under ESO programme ID 179.A-2005 and on data products produced by TERAPIX and the Cambridge Astronomy Survey Unit on behalf of the UltraVISTA consortium. The DEIMOS spectrum presented herein was obtained at the W. M. Keck Observatory, which is operated as a scientific partnership among the California Institute of Technology, the University of California and the National Aeronautics and Space Administration. The Observatory was made possible by the generous financial support of the W. M. Keck Foundation. 
This research made use of Astropy, a community-developed core Python package for Astronomy \citep{astropy13,astropy18}, and XSPEC \citep{arnaud96}.
\par A.M. and A.C. acknowledge support from the ASI/INAF grant I/037/12/0011/13. R.C.H. acknowledges support from the National Science Foundation through CAREER award 1554584, and from NASA through grant number NNX15AP24G. W.N.B. acknowledges support from Caltech NuSTAR subcontract 44A-1092750.

\vspace{5mm}
\facilities{\nustar, \chandra}

\software{NuSTARDAS, FTOOLS}
\bibliographystyle{aasjournal} 
\bibliography{/Volumes/Maxtor/amasini} 

\end{document}